\documentclass[pra,10pt,showpacs,amsmath,twocolumn,floatfix,eqsecnum,superscriptaddress]{revtex4}
\usepackage{amsmath}
\usepackage{amsfonts}
\usepackage{graphicx}

\newcommand{\est}[1]{#1'}
\newcommand{\smooth}[1]{\widetilde{#1}}
\newcommand{\retro}[1]{#1''}

\newcommand{\diff}[2]{\frac{d #1}{d #2}}

\newcommand{\intall}{\int_{-\infty}^{\infty}}
\newcommand{\ket}[1]{|#1\rangle}
\newcommand{\bra}[1]{\langle#1|}

\newcommand{\avg}[1]{\langle#1\rangle}
\newcommand{\Avg}[1]{\left\langle#1\right\rangle}

\newcommand{\bs}[1]{\boldsymbol{#1}}

\newcommand{\bk}[1]{\left(#1\right)}
\newcommand{\Bk}[1]{\left[#1\right]}
\newcommand{\BK}[1]{\left\{#1\right\}}

\newcommand{\trace}{\operatorname{Tr}}
\begin{document}
\title{Quantum theory of optical temporal phase and instantaneous
  frequency. II. Continuous time limit
  and state-variable approach to phase-locked loop design}

\author{Mankei Tsang}

\email{mankei@mit.edu}

\author{Jeffrey H.\ Shapiro}

\affiliation{Research Laboratory of Electronics, Massachusetts
  Institute of Technology, Cambridge, Massachusetts 02139, USA}

\author{Seth Lloyd}

\affiliation{Research Laboratory of Electronics,
Massachusetts Institute of Technology, Cambridge, Massachusetts
02139, USA}

\affiliation{Department of Mechanical Engineering,
Massachusetts Institute of Technology, Cambridge, Massachusetts
02139, USA}

\date{\today}

\begin{abstract}
  We consider the continuous-time version of our recently proposed
  quantum theory of optical temporal phase and instantaneous frequency
  [Tsang, Shapiro, and Lloyd, Phys.\ Rev.\ A \textbf{78}, 053820
  (2008)].  Using a state-variable approach to estimation, we design
  homodyne phase-locked loops that can measure the temporal phase with
  quantum-limited accuracy.  We show that post-processing can further
  improve the estimation performance, if delay is allowed in the
  estimation.  We also investigate the fundamental uncertainties in
  the simultaneous estimation of harmonic-oscillator position and
  momentum via continuous optical phase measurements from the
  classical estimation theory perspective. In the case of delayed
  estimation, we find that the inferred uncertainty product can drop
  below that allowed by the Heisenberg uncertainty relation.  Although
  this result seems counter-intuitive, we argue that it does not
  violate any basic principle of quantum mechanics.
\end{abstract}

\maketitle
\section{Introduction}
Optical phase measurements at the fundamental quantum limit of
accuracy are an important goal in science and engineering and crucial
for future metrology, sensing, and communication applications. While
the single-mode case has been extensively studied, less attention has
been given to the quantum measurements of a temporally varying
phase. Theoretically, the temporal-phase positive operator-valued
measure (POVM) describes the optimal quantum measurements
\cite{tsang}, but it is difficult to perform such measurements in
practice. Adaptive homodyne detection \cite{wiseman,armen} is a much
more feasible approach, and Berry and Wiseman have proposed the use of
a homodyne phase-locked loop to estimate the phase when the mean phase
is a classical Wiener random process \cite{berry,pope}.  On the other
hand, we have recently shown in Ref.~\cite{tsang} how homodyne
phase-locked loops can be designed using classical estimation theory
to perform quantum-limited temporal phase measurements when the mean
phase is any stationary Gaussian random process.

The main purpose of this paper is to unify and generalize the two
distinct approaches undertaken by Berry and Wiseman and ourselves,
under the common framework of classical estimation theory. In
Sec.~\ref{continuous}, we first extend our discrete-time theory
proposed in Ref.~\cite{tsang} to the continuous time domain. In
Sec.~\ref{kb}, we generalize Berry and Wiseman's results to a much
wider class of random processes using the Kalman-Bucy filtering theory
\cite{vantrees,baggeroer}. The Kalman-Bucy approach guarantees the
real-time estimation efficiency provided that the phase-locked loop
operates in the linear regime. Our approach also significantly
simplifies the design of phase-locked loops, compared to the more
computationally expensive Bayesian state estimation approach suggested
by Pope \textit{et al.}\ \cite{pope}. In Sec.~\ref{wiener_loop}, we
show that the Wiener filtering technique used in our previous paper
\cite{tsang} is equivalent to Kalman-Bucy filtering at steady
state. In Sec.~\ref{smooth}, we point out that Berry and Wiseman's
results are not optimal if delay is permitted in the phase estimation
process, and post-processing can further improve the phase estimation
performance beyond that offered by Kalman-Bucy or Wiener filtering. We
illustrate these concepts by considering the specific cases of the
mean phase being an Ornstein-Uhlenbeck random process as well as the
Wiener process studied by Berry and Wiseman. Apart from the
theoretical importance of our results in the context of quantum
estimation and control theory, they should also be of immediate
interest to experimentalists and engineers who wish to achieve
quantum-limited temporal phase measurements, as we expect our
proposals to be realizable using current technology.

In Sec.~\ref{ho_estimation}, we investigate the fundamental problem of
simultaneous harmonic-oscillator position and momentum estimation at
the quantum limit by continuous optical phase measurements. The
problem can be cast directly in the framework of classical estimation
theory for Gaussian states. The use of Kalman-Bucy filtering for
real-time position and momentum estimation has been proposed by
Belavkin and Staszewski \cite{belavkin} and Doherty \textit{et al.}\
\cite{doherty}, who have shown that the quantum state of the harmonic
oscillator conditioned upon the real-time measurement record is a pure
Gaussian state. Here we show that the inferred position and momentum
estimation errors according to classical estimation theory can be
further reduced below the Heisenberg uncertainty product, if delay is
allowed in the estimation. While counter-intuitive, we explain in
Sec.~\ref{discuss} why this result does not violate the basic
principles of quantum mechanics.

\section{\label{continuous}Phase in the continuous time domain}
\begin{figure}[htbp]
\centerline{\includegraphics[width=0.45\textwidth]{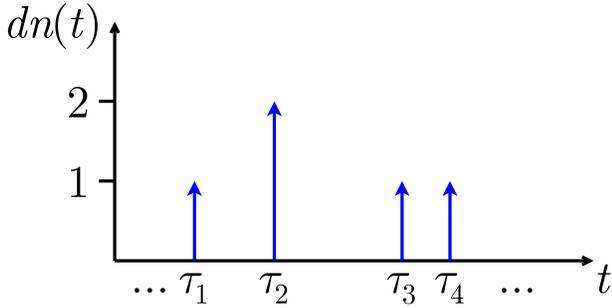}}
\caption{(Color online). A realization of the continuous-time
  discrete-photon-number random process.}
\label{point_process}
\end{figure}

For completeness, we first review the continuous time limit of our
discrete-time theory of temporal phase \cite{tsang}, as previously
described in Ref.~\cite{tsang_qcmc}.  Consider the optical envelope
annihilation and creation operators $\hat{A}(t)$ and
$\hat{A}^\dagger(t)$, respectively, in the slowly varying envelope
regime, with the time-domain commutation relation
\begin{align}
\Bk{\hat{A}(t),\hat{A}^\dagger(t')} &= \delta(t-t').
\end{align}
Let $dn(t)$ be a continuous-time discrete-photon-number random
process, a realization of which is depicted in Fig.~\ref{point_process},
and $\tau_j$ be the times at which $dn(\tau_j)$ is non-zero.

A Fock state with a definite $dn(t)$ can be defined as
\begin{align}
\ket{dn(t)} &\equiv \BK{\prod_j
\frac{1}{\sqrt{dn(\tau_j)!}}\Bk{\hat{A}^\dagger(\tau_j)\sqrt{dt}}^{dn(\tau_j)}}
\ket{0},
\end{align}
which is an eigenstate of the photon-number flux operator
$\hat{A}^\dagger(t)\hat{A}(t)$,
\begin{align}
\hat{A}^\dagger(t)\hat{A}(t)\ket{dn(t)} &= 
I(t)\ket{dn(t)},
\\
I(t) \equiv \frac{dn(t)}{dt} &= \sum_j dn(\tau_j)\delta(t-\tau_j).
\end{align}
The Fock states form a complete orthogonal basis of the
continuous-time Hilbert space,
\begin{align}
\sum_{dn(t)}\ket{dn(t)}\bra{dn(t)} &= \hat{1},
\end{align}
where the sum is over all realizations of $dn(t)$.
For a quantum state $\hat\rho$, the photon-number probability
distribution is
\begin{align}
P[dn(t)] &= \trace\BK{\hat\rho\ket{dn(t)}\bra{dn(t)}},
&
\sum_{dn(t)}P[dn(t)] &= 1.
\end{align}
For example, a coherent state is defined as
\begin{align}
\ket{\mathcal{A}(t)} &= 
\exp\Bk{-\frac{\bar N}{2} + \intall dt \mathcal{A}(t)\hat{A}^\dagger(t)}
\ket{0},
\nonumber\\
\bar N &\equiv \intall dt |\mathcal{A}(t)|^2,
\quad
\hat{A}(t)\ket{\mathcal{A}(t)}=\mathcal{A}(t)\ket{\mathcal{A}(t)},
\end{align}
where $\mathcal A(t)$ is the mean field. The photon-number probability
density is then
\begin{align}
P[dn(t)] &= \lim_{\delta t \to dt}e^{-\bar N}\prod_j
\frac{\Bk{|\mathcal{A}(t_j)|^2 \delta t}^{dn(t_j)}}{dn(t_j)!},
\nonumber\\
t_j  &\equiv t_0 + j\delta t,
\end{align}
which describes a Poisson process, as is well known
\cite{hudson}.

A temporal phase state can be defined
as the functional Fourier transform of the Fock states,
\begin{align}
\ket{\phi(t)} &\equiv \sum_{dn(t)}
\exp\Bk{i\sum_j dn(\tau_j)\phi(\tau_j)} \ket{dn(t)}
\nonumber\\
&=
\sum_{dn(t)}
\exp\Bk{i\intall dt I(t)\phi(t)} \ket{dn(t)}.
\end{align}
In terms of the temporal phase states, a temporal-phase
POVM can be defined as
\begin{align}
\hat{\Pi}\Bk{\phi(t)} &\equiv \ket{\phi(t)}\bra{\phi(t)},
\end{align}
which is the continuous limit of the one defined in Ref.~\cite{tsang}
and can be normalized using a path integral with the
paths restricted to a range of $2\pi$,
\begin{align}
\int D\phi(t) \hat{\Pi}\Bk{\phi(t)} &= \hat{1},
\quad
D\phi(t) \equiv \lim_{\delta t \to dt}\prod_j \frac{d\phi(t_j)}{2\pi},
\nonumber\\
\phi_0(t) &\le \phi(t) < \phi_0(t)+2\pi.
\end{align}
The temporal-phase probability density is thus given by
\begin{align}
p[\phi(t)] &= \trace\BK{\hat\rho\hat\Pi[\phi(t)]},
&
\int D\phi(t)p[\phi(t)] &= 1.
\end{align}
It is difficult to analytically calculate $p[\phi(t)]$ for most
quantum states of interest, so perturbative or numerical methods
should be sought.

For the design of homodyne phase-locked loops, the Wigner distribution
is of more interest. For a Gaussian state with uncorrelated
quadratures, it can be written as
\begin{align}
&\quad W[\xi_1(t),\xi_2(t)]
\nonumber\\
&\propto \exp\Bk{-\frac{1}{2}\sum_{j = 1,2}
\int dt d\tau \xi_j(t)K_{j}^{-1}(t,\tau)\xi_j(\tau)},
\end{align}
where $\xi_j(t)$ are quadrature processes,
\begin{align}
\xi_1(t) &\equiv A(t)e^{-i\theta(t)}+
A^*(t)e^{i\theta(t)}
\nonumber\\&\quad-
\Avg{A(t)e^{-i\theta(t)}+A^*(t)e^{i\theta(t)}},
\\
\xi_2(t) &\equiv -i\Bk{
A(t)e^{-i\theta(t)}-A^*(t)e^{i\theta(t)}}
\nonumber\\&\quad-
\Avg{-i\Bk{A(t)e^{-i\theta(t)}-A^*(t)e^{i\theta(t)}}},
\end{align}
$A(t)$ is the complex field variable in phase space, $\theta$ is an
arbitrary phase, and $K_j^{-1}(t,\tau)$ is defined in terms of the
covariance functions $K_j(t,\tau) $ as
\begin{align}
\int du K_j(t,u)K_j^{-1}(u,\tau) &= \delta(t-\tau),
\\
K_j(t,\tau) &\equiv \Avg{\xi_j(t)\xi_j(\tau)}.
\end{align}
The covariance functions must satisfy the uncertainty relation
\begin{align}
\int du K_1(t,u)K_2(u,\tau) &\ge \delta(t-\tau),
\end{align}
which becomes an equality if and only if the state is pure. In
particular, the covariance functions for a coherent state are
\begin{align}
K_1(t,\tau) &= K_2(t,\tau) = \delta(t-\tau).
\end{align}

\section{Phase-locked loop design}
\subsection{\label{kb}Kalman-Bucy filtering}
\begin{figure}[htbp]
\centerline{\includegraphics[width=0.45\textwidth]{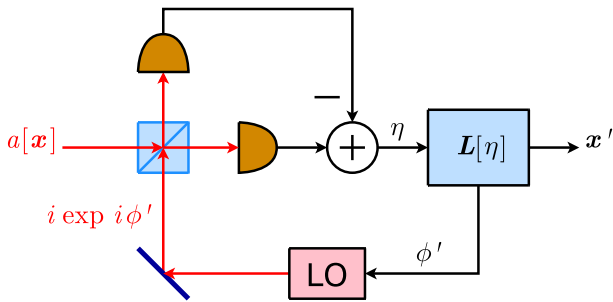}}
\caption{(Color online). A homodyne phase-locked loop. LO denotes
  local oscillator.}
\label{homodyne_pll}
\end{figure}
Consider the homodyne phase-locked loop illustrated in
Fig.~\ref{homodyne_pll}.  The output of the homodyne detection can be
written as
\begin{align}
\eta(t) &= \sin[\bar\phi(t)-\est{\phi}(t)] + z(t),
\label{homodyne}
\end{align}
where $\bar\phi(t)$ is the mean phase of the optical field, which
contains the message to be estimated, $\est{\phi}(t)$ is the
local-oscillator phase, and $z(t)$ is the quantum noise. For a
phase-squeezed state with squeezed quadrature $\xi_2(t)$ and
anti-squeezed quadrature $\xi_1(t)$, $z(t)$ can be written as
\begin{align}
z(t) &\equiv \frac{1}{2|\mathcal{A}|}
\Big\{\xi_1(t)\sin[\bar\phi(t)-\est\phi(t)]
\nonumber\\&\quad
+\xi_2(t)\cos[\bar\phi(t)-\est\phi(t)]\Big\},
\end{align}
where $\mathcal A \equiv \avg{\hat A} = |\mathcal A|\exp(i\bar\phi)$
is the mean field.  For generality, we let the message be a vector of
$n$ random processes,
\begin{align}
\bs{x}(t) &\equiv \Bk{\begin{array}{c}x_1(t)\\
x_2(t)\\\vdots\\x_n(t)\end{array}},
\end{align}
with the mean phase proportional to the first one,
\begin{align}
\bar\phi(t) &= \bs{C}(t) \bs{x}(t),
&
\bs{C}(t) &\equiv [\beta, 0,\dots,0].
\end{align}
In the Kalman-Bucy formalism, $\bs{x}(t)$ is modeled as zero-mean
random processes that satisfy a set of linear differential equations,
\begin{align}
\diff{\bs{x}(t)}{t} &= \bs{A}(t)\bs{x}(t) + \bs{B}(t)\bs{u}(t),
\label{evolution}
\end{align}
where $\bs{A}(t)$ and $\bs{B}(t)$ are $n\times n$ and $n\times m$
matrices, respectively, and $\bs{u}(t)$ is a vector of $m$ zero-mean
white Gaussian inputs with autocorrelation
\begin{align}
\Avg{\bs{u}(t)\otimes\bs{u}(\tau)} &= \bs{U}\delta(t-\tau).
\end{align}
We focus on coherent states, so that $z(t)$ can be modeled as an
independent white Gaussian noise according to its Wigner distribution,
\begin{align}
\Avg{z(t)z(\tau)} &= Z(t)\delta(t-\tau),
&
Z(t) &\equiv \frac{1}{4|\mathcal{A}|^2} = 
\frac{\hbar\omega_0}{4\mathcal{P}}.
\end{align}
where $\omega_0$ is the optical carrier frequency and $\mathcal{P}$ is
the average optical power. The additive white Gaussian noise allows us
to apply classical estimation techniques directly. Coherent states
should also be of more immediate interest to experimentalists and
engineers, as they are easier to generate and more robust to loss
compared to nonclassical states.  For a phase-squeezed state, the
statistics of $z(t)$ depend on $\bar\phi(t)-\est{\phi}(t)$, but one
may still wish to approximate $z(t)$ as an independent Gaussian noise
by neglecting the anti-squeezed quadrature $\xi_1(t)$, in order
to take advantage of classical estimation techniques \cite{tsang}.

The purpose of the phase-locked loop is to make $\est\phi(t)$ the
optimal estimate of $\bar\phi(t)$, using the
measurement record of $\eta(\tau)$ in the period $t_0 \le \tau \le t$,
such that we can linearize Eq.~(\ref{homodyne}),
\begin{align}
\eta(t) &\approx \bar\phi(t)-\est{\phi}(t) + z(t),
\end{align}
when the following condition, called the threshold constraint
in classical estimation theory \cite{tsang,vantrees}, is satisfied,
\begin{align}
\avg{[\bar\phi(t)-\est\phi(t)]^2} \ll 1.
\label{threshold}
\end{align}
The threshold constraint ensures that the phase-locked loop is phase-locked.

If the canonical measurements characterized by the temporal-phase POVM
can be performed, we can instead modulate the phase of the incoming
field by $-\est\phi(t)$ and perform the canonical measurements,
producing an output
\begin{align}
\eta_c(t) &= f\big(\bar\phi(t)-\est\phi(t)+ z(t)\big),
\end{align}
where $f(\phi)$ must be a periodic function, such as a sawtooth
function,
\begin{align}
f(\phi) &= \Bk{\bk{\phi-\pi} {\rm mod}\ 2\pi} - \pi,
\end{align}
and $z(t)$ is the quantum phase noise and independent of $\bar\phi(t)$
and $\est\phi(t)$ for any quantum state. Because $\bar\phi(t)$ may
exceed the $2\pi$ range, it is still necessary to use the phase-locked
loop to perform phase unwrapping. The following analysis can be
applied to canonical temporal-phase measurements and arbitrary quantum
states if $\eta_{c}(t)$ is linearized as
\begin{align}
\eta_{c}(t) &\approx \bar\phi(t)-\est\phi(t)+ z(t),
\end{align}
and $z(t)$ is approximated as a white Gaussian noise. The same
threshold constraint given by Eq.~(\ref{threshold}) ensures that the
periodic nature of $\eta_{c}(t)$ can be neglected and the
linearization is valid. 

\begin{figure}[htbp]
\centerline{\includegraphics[width=0.45\textwidth]{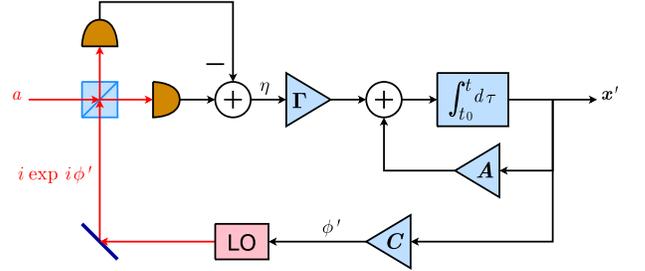}}
\caption{(Color online). A phase-locked loop that implements
  Kalman-Bucy filtering for angle demodulation.}
\label{phase_estimator}
\end{figure}

The linearization allows us to use Kalman-Bucy filtering to produce
the real-time minimum-mean-square-error estimates of $\bs{x}(t)$
\cite{vantrees,baggeroer}, which we denote as $\est{\bs{x}}(t)$,
\begin{align}
\diff{\est{\bs{x}}}{t} &= \bs{A}\est{\bs{x}} +
\bs{\Gamma}\bs{\eta}.
\label{estimator}
\end{align}
This is called the Kalman-Bucy \emph{estimator equation}. $\bs{\eta} $
is called the \emph{innovation}, defined in terms of a general
vectoral observation process $\bs y(t)$ as
\begin{align}
\bs y(t) &\equiv \bs C(t)\bs x(t) + \bs z(t),
&
\bs \eta(t) &\equiv \bs y(t)-\bs C(t)\bs x'(t),
\end{align}
where $\bs z(t)$ is a vectoral Gaussian white noise with mean
$\avg{\bs z(t)} = \bs 0$ and covariance $\avg{\bs z(t)\otimes\bs
  z(\tau)}\equiv \bs Z(t)\delta(t-\tau)$.  For phase-locked loops, the
homodyne output $\eta(t)$ can be used directly as the innovation, so
$\bs z(t) = z(t)$ and $\bs Z(t) = Z(t)$. $\bs{\Gamma}$ is called the
\emph{gain}, given by
\begin{align}
\bs\Gamma &= \bs{\Sigma}\bs{C}^T \bs{Z}^{-1}
 = \frac{4\beta\mathcal{P}}{\hbar\omega_0}
\Bk{\begin{array}{c}\Sigma_{11}\\\Sigma_{21}\\
\vdots\\\Sigma_{n1}\end{array}},
\end{align}
and $\bs{\Sigma}$ is the estimation covariance matrix, defined as
\begin{align}
\bs{\Sigma}(t) &\equiv \Avg{[\bs{x}(t)-\est{\bs{x}}(t)]
\otimes[\bs{x}(t)-\est{\bs{x}}(t)]},
\end{align}
which satisfies the \emph{variance equation},
\begin{align}
\diff{\bs{\Sigma}}{t} &= \bs{A}\bs{\Sigma}
+\bs{\Sigma}\bs{A}^T 
-\bs{\Sigma}\bs{C}^T\bs{Z}^{-1}\bs{C}\bs{\Sigma}
+\bs{B}\bs{U}\bs{B}^T.
\label{variance_eq}
\end{align}
Equations (\ref{estimator}) to (\ref{variance_eq}) are much simpler to
solve than the conditional probability density equation suggested by
Pope \textit{et al.}\ for phase estimation \cite{pope}.  The threshold
constraint becomes
\begin{align}
\beta^2\Sigma_{11} \ll 1,
\end{align}
and the initial conditions are
\begin{align}
\est{\bs{x}}(t_0) &= \avg{\bs{x}(t_0)} = \bs{0},
&
\bs{\Sigma}(t_0) &= \avg{\bs{x}(t_0)\otimes\bs{x}(t_0)}.
\end{align}
Apart from phase estimation, Kalman-Bucy filtering can also be used to
simultaneously estimate other parameters that depend linearly on the
phase. The instantaneous frequency, for instance, can be estimated by
defining $x_2 \propto dx_1/dt$. The phase-locked-loop implementation
of Kalman-Bucy filtering for general angle demodulation is depicted in
Fig.~\ref{phase_estimator}.

For example, consider the message as an Ornstein-Uhlenbeck process,
\begin{align}
\diff{x}{t} &= -k x + B u.
\end{align}
The variance equation becomes
\begin{align}
\diff{\Sigma}{t} &= -2k\Sigma
-\frac{4\beta^2\mathcal{P}}{\hbar\omega_0}\Sigma^2
+\kappa, &
\kappa &\equiv B^2 U,
\end{align}
and the gain is
\begin{align}
\Gamma(t) &= \frac{4\beta\mathcal{P}}{\hbar\omega_0}\Sigma(t).
\end{align}
The variance equation can be solved analytically,
\begin{align}
\Sigma(t) &= \Sigma_{\rm ss} \frac{\mu-
\frac{\gamma/k+1}{\gamma/k-1}\exp[-2\gamma (t-t_0)]}
{\mu+\exp[-2\gamma (t-t_0)]},
\\
\mu &\equiv \frac{\gamma/k+1+\Lambda\Sigma(t_0)}
{\gamma/k-1-\Lambda\Sigma(t_0)},
\quad
\gamma \equiv k\bk{\frac{\kappa\Lambda}{k}+1}^{1/2},
\nonumber\\
\Lambda &\equiv \frac{4\beta^2\mathcal{P}}{\hbar\omega_0k}.
\label{Ornstein-Uhlenbeck_parameters}
\end{align}
where the subscript ss denotes the steady state,
\begin{align}
\Sigma(t) &\to \Sigma_{\rm ss}\equiv 
\frac{1}{\Lambda}\Bk{\bk{\frac{\kappa\Lambda}{k}+1}^{1/2}-1},
\quad t-t_0 \gg \frac{1}{\gamma},
\\
\Sigma_{\rm ss}&\approx \bk{\frac{\kappa}{k\Lambda}}^{1/2}, 
\quad \frac{\kappa\Lambda}{k} \gg 1,
\end{align}
and the threshold constraint is
\begin{align}
\Lambda \gg \frac{\kappa}{k}\beta^4.
\end{align}
When the message is a Wiener random process, 
\begin{align}
\diff{x}{t} &= Bu,
\end{align}
we can either follow the same procedure as before to derive
the Kalman-Bucy filter, or take the results for the Ornstein-Uhlenbeck
process to the limit $k \to 0$. Either way, assuming
$\beta = 1$ for simplicity, we find
\begin{align}
\Sigma(t) &= \Sigma_{\rm ss}
\frac{\mu-\exp[-2\gamma(t-t_0)]}{\mu+\exp[-2\gamma(t-t_0)]},
\quad
\Gamma(t) = \frac{4\mathcal{P}}{\hbar\omega_0}\Sigma(t),
\\
\Sigma_{\rm ss} &= \frac{1}{2\sqrt{N}},
\quad
\mu \equiv \frac{1+2\sqrt{N}\Sigma(t_0)}{1-2\sqrt{N}\Sigma(t_0)},
\nonumber\\
\gamma &\equiv 2\kappa\sqrt{N},
\quad
N \equiv \frac{\mathcal{P}}{\hbar\omega_0\kappa}.
\end{align}
At steady state,
\begin{align}
\Sigma(t)  \to\Sigma_{\rm ss}
&=\frac{1}{2\sqrt{N}},
&
\Gamma(t) &\to 2\kappa\sqrt{N},
&
t-t_0 \gg \frac{1}{\gamma}.
\end{align}
The threshold constraint is
\begin{align}
4N = \frac{4\mathcal{P}}{\hbar\omega_0\kappa} \gg 1.
\end{align}
These results for the Wiener process agree with Berry and Wiseman's \cite{berry}.

\subsection{\label{wiener_loop}Wiener filtering}
\begin{figure}[htbp]
\centerline{\includegraphics[width=0.45\textwidth]{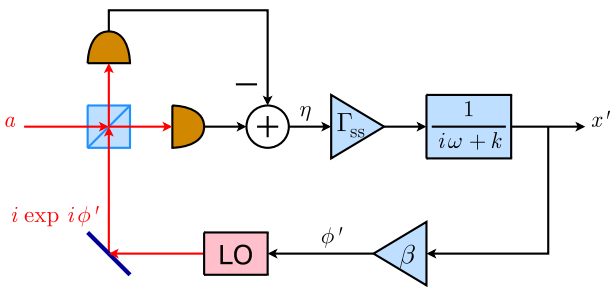}}
\caption{(Color online). A phase-locked loop implementation of Wiener
  filtering when the mean phase is an Ornstein-Uhlenbeck process.}
\label{wiener}
\end{figure}

In addition to the Kalman-Bucy state-variable approach, Wiener's
frequency-domain approach can also be used to design the phase-locked
loop \cite{tsang,vantrees,viterbi}. Defining
\begin{align}
y(t) &\equiv \bar\phi(t) + z(t),
\end{align}
it can be shown that Kalman-Bucy filtering is equivalent to the integral
equation \cite{vantrees,baggeroer}
\begin{align}
\est{\bs{x}}(t) &= \int_{t_0}^t d\tau \bs{H}(t,\tau)y(\tau),
\end{align}
where $\bs{H}(t,\tau)$ is called the optimum realizable filter
and satisfies the integral equation
\begin{align}
\bs{K}_{xy}(t,\sigma) &= \int_{t_0}^t d\tau\bs{H}(t,\tau)
K_y(\tau,\sigma),
\label{integral}\\
\bs{K}_{xy}(t,\sigma) &\equiv \Avg{\bs{x}(t) y(\sigma)},
\quad
K_y(\tau,\sigma) = \Avg{y(\tau)y(\sigma)}.
\end{align}
If $\bs{x}(t)$ and $y(t)$ are stationary and we let $t_0\to-\infty$,
Eq.~(\ref{integral}) becomes the Wiener-Hopf equation,
\begin{align}
\bs{K}_{xy}(t-\sigma) &=\int_{-\infty}^t d\tau
\bs{H}(t-\tau)K_y(\tau-\sigma),
\end{align}
which can be solved by a well-known frequency-domain
technique \cite{tsang,vantrees,viterbi}.
For example, if $x(t)$ is an Ornstein-Uhlenbeck process,
its power spectral density in the limit of
$t_0\to-\infty$ is
\begin{align}
K_x(t,\sigma) &\equiv \Avg{x(t)x(\sigma)} = K_x(t-\sigma),
\\
S_x(\omega) &\equiv \intall dt K_x(t)\exp(-i\omega t)
=\frac{\kappa}{\omega^2+k^2}.
\end{align}
The power spectral density for $y(t)$ is then
\begin{align}
S_y(\omega) &\equiv \intall dt K_y(t)\exp(-i\omega t)
=\frac{\beta^2\kappa}{\omega^2+k^2} + \frac{\hbar\omega_0}{4\mathcal{P}}.
\end{align}
To solve for $H(t-\tau)$, we rewrite $S_y(\omega)$  as
\begin{align}
S_y(\omega) &= H_+(\omega)H_+^*(\omega),
&
H_+(\omega) &= \bk{\frac{\hbar\omega_0}{4\mathcal{P}}}^{1/2}
\frac{i\omega+\gamma}{i\omega + k},
\end{align}
where $\gamma$ is given in Eq.~(\ref{Ornstein-Uhlenbeck_parameters}),
and $H_+(\omega)$ and $1/H_+(\omega)$ are causal filters.
Defining
\begin{align}
S_{xy}(\omega) &\equiv \intall dt K_{xy}(t)\exp(-i\omega t)
=\frac{\beta\kappa}{\omega^2+k^2},
\end{align}
the Wiener filter in the frequency domain is
\begin{align}
H(\omega) &= \frac{1}{H_+(\omega)}
\Bk{\frac{S_{xy}(\omega)}{H_+^*(\omega)}}_+,
\end{align}
where the subscript $+$ denotes the realizable part. To calculate
the realizable part, first perform the inverse Fourier transform,
\begin{align}
&\quad\int \frac{d\omega}{2\pi}\frac{S_{xy}(\omega)}{H_+^*(\omega)}
\exp(i\omega t)
\nonumber\\
&= \bk{\frac{4\mathcal P}{\hbar\omega_0}}^{1/2}
\frac{\beta\kappa}{\gamma+k}
\Bk{\mathcal U(-t)e^{\gamma t}+ \mathcal U(t)e^{-kt}},
\label{ift}
\end{align}
where $\mathcal U(t)$ is the Heaviside step function. The realizable
part is then obtained by multiplying Eq.~(\ref{ift}) by $\mathcal
U(t)$ and performing the Fourier transform.  After some algebra, we
obtain
\begin{align}
H(\omega) &= \frac{\Gamma_{\rm ss}}{i\omega + \gamma},
&
\Gamma_{\rm ss} &\equiv \frac{\gamma-k}{\beta}.
\end{align}
To implement the Wiener filter in the phase-locked loop shown in
Fig.~\ref{wiener}, the loop filter that relates the homodyne output
$\eta(t)$ to the estimate $\est{x}(t)$ is
\begin{align}
\est{x}(t) &= \int_{-\infty}^t d\tau L(t-\tau)\eta(\tau)
\\
&\approx 
\int_{-\infty}^t d\tau L(t-\tau)[y(\tau)-\beta\est{x}(\tau)],
\\
\frac{L(\omega)}{1+\beta L(\omega)} &= H(\omega),
\quad
L(\omega) = \frac{H(\omega)}{1-\beta H(\omega)} = 
\frac{\Gamma_{\rm ss}}{i\omega + k}.
\end{align}
The resulting phase-locked-loop structure is equivalent to that
obtained by Kalman-Bucy filtering at steady state, as both approaches
implement the optimum realizable filter.  

The mean-square error of Wiener filtering is given by
the well-known expression
\cite{tsang,vantrees,viterbi}
\begin{align}
\Sigma &= \frac{\hbar\omega_0}{4\beta^2\mathcal{P}}
\intall \frac{d\omega}{2\pi}
 \ln \Bk{1+\frac{4\beta^2\mathcal{P}S_x(\omega)}{\hbar\omega_0}}
\nonumber\\
&=\frac{1}{\Lambda}\Bk{\bk{\frac{\kappa\Lambda}{k}+1}^{1/2}-1},
\end{align}
which obviously must be the same as the steady-state error
$\Sigma_{\rm ss}$ obtained by Kalman-Bucy filtering. The interested
reader is referred to Ref.~\cite{vantrees} for an excellent treatment
of Wiener filters.

The advantage of Kalman-Bucy filtering over Wiener filtering is that
the former can also deal with a wide class of nonstationary random
processes that can be described by a system of linear equations
(\ref{evolution}), whereas Wiener filtering works only for stationary
processes.  In the special case of a Wiener process, however, we can
first design a Wiener filter for an Ornstein-Uhlenbeck process and
take the limit $k\to 0$. The result for $\beta = 1$ is
\begin{align}
L(\omega) &\to \frac{\Gamma_{\rm ss}}{i\omega},
&
\Gamma_{\rm ss} &\to 2\kappa\sqrt{N},
&
\Sigma &\to \frac{1}{2\sqrt{N}},
\end{align}
which is again the same as the steady-state Kalman-Bucy filter.

\section{\label{smooth}Smoothing}
Both Kalman-Bucy filtering and Wiener filtering provide real-time estimates of
$\bs{x}(t)$ based on the measurement record up to time $\tau = t$. If
we allow delay in the estimation, we can use the additional
information from more advanced measurements to improve upon the
estimation. In the following we consider the optimal estimation of
$\bs{x}(t)$ given the full measurement record in the interval $t_0 \le
t \le T$, also called smoothing in classical estimation theory
\cite{vantrees,baggeroer}.

\subsection{\label{rts}State-variable approach}
Given the output $\est{\bs x}$ of the homodyne phase-locked loop
designed by Kalman-Bucy filtering and the associated covariance matrix
$\bs\Sigma$, the optimal smoothing estimates of $\bs{x}(t)$, which we
define as $\smooth{\bs x}(t)$, can be calculated using a
state-variable approach, first suggested by Bryson and Frazier
\cite{baggeroer,rts}.  $\smooth{\bs{x}}(t)$ and the smoothing
covariance matrix,
\begin{align}
\bs{\Pi}(t) &\equiv \Avg{\Bk{\bs{x}(t)-\smooth{\bs x}(t)}
\otimes\Bk{\bs{x}(t)-\smooth{\bs x}(t)}},
\end{align}
 can be obtained by solving the following
equations backward in time,
\begin{align}
\diff{\smooth{\bs{x}}}{t} &= \bs{A}\smooth{\bs x} + 
\bs{B}\bs{U}\bs{B}^T\bs\Sigma^{-1}
\bk{\smooth{\bs{x}}-\est{\bs x}},
\label{rts_eqn}
\\
\diff{\bs{\Pi}}{t} &= 
\bk{\bs{A}+\bs{B}\bs{U}\bs{B}^T\bs\Sigma^{-1}}\bs{\Pi}
 + 
\bs{\Pi}\bk{\bs{A}+\bs{B}\bs{U}\bs{B}^T\bs\Sigma^{-1}}^T
\nonumber\\&\quad
-\bs{B}\bs{U}\bs{B}^T,
\label{rts_var}
\end{align}
with the final conditions,
\begin{align}
\smooth{\bs{x}}(T) &= \est{\bs x}(T),
&
\bs{\Pi}(T) &= \bs\Sigma(T).
\end{align}
In the $t_0 \ll t \ll T$ limit, we can calculate the
steady-state smoothing covariance matrix $\bs{\Pi}_{\rm ss}$ by
setting the right-hand side Eq.~(\ref{rts_var}) to zero and
using the steady-state $\bs\Sigma_{\rm ss}$ as $\bs\Sigma$.

Again using the Ornstein-Uhlenbeck process as an example, the
steady-state smoothing error, also called the ``irreducible'' error
\cite{vantrees,viterbi}, is given by
\begin{align}
\Pi_{\rm ss} &=
\frac{\kappa}{2k(\kappa\Lambda/k+1)^{1/2}}.
\label{Pss}
\end{align}
This result is identical to that derived in \cite{tsang} using a
frequency-domain approach.  In the limit of $\Lambda \gg k/\kappa$,
\begin{align}
\Pi_{\rm ss} &\to \frac{1}{2}\bk{\frac{\kappa}{k\Lambda}}^{1/2} \approx
\frac{1}{2}\Sigma_{\rm ss},
\end{align}
which is smaller than the error from Kalman-Bucy or Wiener filtering by
approximately a factor of 2. For the Wiener process, 
the smoothing error is
\begin{align}
\Pi_{\rm ss} &= \frac{1}{4\sqrt{N}} = \frac{1}{2}\Sigma_{\rm ss},
\end{align}
which is smaller than the filtering error by exactly a factor of 2.

\subsection{\label{twofilter}Two-filter smoothing}
An equivalent but more intuitive form of the optimal smoother
was discovered by Mayne \cite{mayne} and Fraser and Potter
\cite{fraser}, who treat the smoother as a combination of two filters,
one running forward in time to produce a prediction $\est{\bs{x}}(t)$
via Kalman-Bucy filtering using the past measurement record, as
specified by Eqs.~(\ref{estimator}) to (\ref{variance_eq}), and one
running backward in time to produce a \emph{retrodiction}
$\bs{x}''(t)$ using the advanced measurement record,
\begin{align}
\diff{\retro{\bs x}}{t} &=
\bs{A}\retro{\bs x} -\bs\Upsilon\bs{\eta},
\\
\diff{\bs\Xi}{t} &= 
\bs{A}\bs\Xi
+\bs\Xi\bs{A}^T 
+\bs\Xi\bs{C}^T\bs Z^{-1}\bs{C}\bs\Xi
-\bs{B}\bs{U}\bs{B}^T,
\\
\bs\Upsilon &= \bs\Xi\bs{C}^T \bs Z^{-1},
\end{align}
with final conditions
\begin{align}
\bs\Xi^{-1}(T)\retro{\bs x}(T) &= \bs{0}, 
& \bs\Xi^{-1}(T) = \bs 0.
\end{align}
The smoothing estimates and covariance matrix, taking into account
both the prediction and the retrodiction, are given by
\begin{align}
\smooth{\bs x} &= \bs{\Pi}
\bk{\bs\Sigma^{-1}\est{\bs x} + \bs\Xi^{-1}\retro{\bs x}},
\\
\bs{\Pi} &= \bk{\bs\Sigma^{-1}+\bs\Xi^{-1}}^{-1}.
\end{align}
The steady-state smoothing covariance matrix $\bs{\Pi}_{\rm ss}$ can be
calculated by combining the steady-state predictive and retrodictive
covariance matrices,
\begin{align}
\bs{\Pi}_{\rm ss} &= \bk{\bs\Sigma^{-1}_{\rm ss}+\bs\Xi^{-1}_{\rm ss}}^{-1}.
\end{align}

\subsection{\label{delay}Frequency-domain approach}
\begin{figure}[htbp]
\centerline{\includegraphics[width=0.45\textwidth]{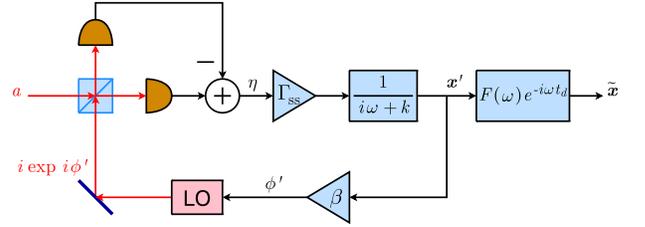}}
\caption{(Color online). A homodyne phase-locked loop with a post-loop
  filter $F(\omega)$ that realizes optimal smoothing.}
\label{delay_estimator}
\end{figure}

For stationary Gaussian random processes and in the limit of $t_0\ll t
\ll T$, a frequency-domain approach can also be used to obtain the
optimal smoother \cite{tsang,vantrees,viterbi}.  The optimal smoothing
estimates can be written in terms of $y(t)$ as
\cite{vantrees,baggeroer,viterbi}
\begin{align}
\smooth{\bs{x}}(t) &= \int_{t_0}^{T} d\tau\bs{G}(t,\tau)y(\tau),
\end{align}
where $\bs{G}(t,\tau)$ obeys
\begin{align}
\bs{K}_{xy}(t,\sigma) &=
\int_{t_0}^{T} d\tau \bs{G}(t,\tau)K_{y}(\tau,\sigma).
\label{optimum_unreal_filter}
\end{align}
For $\bs{K}_{xy}(t,\sigma)=\bs{K}_{xy}(t-\sigma)$,
$\bs{G}(t,\tau) = \bs{G}(t-\tau)$, and
$K_y(\tau,\sigma) = K_y(\tau-\sigma)$,
and in the limit of $t_0\to -\infty$ and $T\to\infty$,
we can solve
Eq.~(\ref{optimum_unreal_filter}) by Fourier transform,
\begin{align}
\bs{G}(\omega)&= \frac{\bs{S}_{xy}(\omega)}{S_{y}(\omega)}.
\end{align}
$\bs{G}(\omega)$ is called the optimum unrealizable filter
\cite{vantrees}.  For an Ornstein-Uhlenbeck process, $G(\omega)$ is
\begin{align}
G(\omega) &= \frac{4\beta\mathcal{P}}{\hbar\omega_0}
\frac{\kappa}{\omega^2+\gamma^2}.
\end{align}
To implement this filter, one can use the homodyne phase-locked loop designed
by Wiener filtering and a post-loop filter given by
\begin{align}
F(\omega) &= \frac{G(\omega)}{H(\omega)}
=\frac{k+\gamma}{-i\omega + \gamma}.
\end{align}
The post-loop filter impulse response is
\begin{align}
f(t) &\equiv \intall \frac{d\omega}{2\pi}
F(\omega)\exp(i\omega t)
\\
&= \Big\{\begin{array}{cc}
(k+\gamma)\exp(\gamma t), & t \le 0,
\\
0, & t > 0,
\end{array}
\end{align}
which is anti-causal, so one must introduce a time delay $t_d \gg
1/\gamma$ for $F$ to be approximated by a causal filter.  The optimal
smoother designed by the frequency-domain approach is depicted
in Fig.~\ref{delay_estimator}.

The variance of the optimal frequency-domain smoother
is \cite{tsang,vantrees,viterbi}
\begin{align}
\Pi &=\intall\frac{d\omega}{2\pi}\Bk{S_x(\omega)-\frac{|S_{xy}(\omega)|^2}
{S_y(\omega)}}
\nonumber\\
&=\frac{\kappa}{2k(\kappa\Lambda/k+1)^{1/2}},
\end{align}
which is the same as the smoothing error derived by the state-variable
approach in Eq.~(\ref{Pss}), as expected.  The interested reader is
again referred to Ref.~\cite{vantrees} for an excellent treatment of
optimal frequency-domain filters and smoothers.

\section{\label{ho_estimation}Quantum position and momentum
estimation by optical phase measurements}
\subsection{Quantum Kalman-Bucy filtering}
So far we have assumed that the mean phase of the optical field
contains classical random processes to be estimated in the presence of
quantum optical noise. In this section we investigate the estimation
of inherently quantum processes carried by the optical
phase. Specifically, we revisit the classic problem of quantum-limited
mirror position and momentum estimation by optical phase measurements.
First we review the problem of optimal real-time estimation by
Kalman-Bucy filtering, previously studied by Belavkin and Staszewski
\cite{belavkin} and Doherty \textit{et al.}\ \cite{doherty}.

\begin{figure}[htbp]
\centerline{\includegraphics[width=0.45\textwidth]{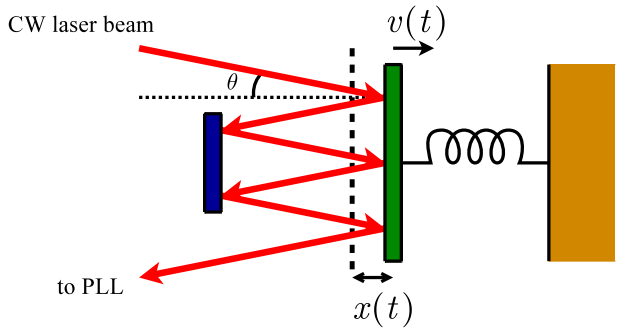}}
\caption{(Color online). Position and momentum estimation by optical
  phase measurements.}
\label{velocimeter}
\end{figure}

We model the mirror as a harmonic oscillator, as depicted in
Fig.~\ref{velocimeter},
\begin{align}
\diff{\hat x}{t} &= \frac{\hat p}{m},
\label{xt}
\\
\diff{\hat{p}}{t} &= -m\omega_m^2 \hat{x}
 + \frac{2 M \hbar\omega_0 \cos\theta }{c}\hat{I},
\label{vt}
\end{align}
where $\hat{x}(t)$ and $\hat{p}(t)$ are quantum position and momentum
operators, $m$ is the harmonic-oscillator mass, $\omega_m$ is the
mechanical harmonic-oscillator frequency, the last term of
Eq.~(\ref{vt}) is the radiation pressure term, $M$ is the number of
times the optical beam hits the mirror, $\theta$ is the angle at which
the optical beam hits the mirror, and $\hat{I}(t)$ is the optical flux
operator, consisting of a mean and a quantum noise term,
\begin{align}
\hat{I}(t) &= \frac{\mathcal{P}}{\hbar\omega_0} + \Delta\hat{I}(t).
\end{align}
The constant force term can be eliminated by redefining
the position of the harmonic oscillator, 
so we shall neglect the constant radiation pressure term from now
on. $\Delta \hat I(t)$ is approximately a white Gaussian noise term for a
high-power optical coherent state,
\begin{align}
\Avg{\Delta\hat I(t)\Delta\hat I(\tau)} &\approx 
\frac{\mathcal{P}}{\hbar\omega_0}\delta(t-\tau).
\end{align}
The Gaussian approximation neglects the discreteness of photon number,
and is valid if the number of photons within the relaxation time of
the filter impulse response is much larger than 1.  We can then write
the quantum system model as
\begin{align}
\diff{}{t}\Bk{\begin{array}{c}
\hat x
\\
\hat p
\end{array}
} &= 
\Bk{\begin{array}{cc}
0 & 1/m
\\
-m\omega_m^2 & 0
\end{array}}
\Bk{\begin{array}{c}
\hat x
\\
\hat p
\end{array}
}
+
\Bk{\begin{array}{c}
0
\\
1
\end{array}
}\hat u,
\end{align}
with initial conditions
\begin{align}
\avg{\hat x(t_0)} &= 0, &\avg{\hat p(t_0)} &= 0,
\label{initial}
\end{align}
and radiation pressure acting as the quantum Langevin noise,
\begin{align}
\Avg{\hat u(t)\hat u(\tau)} = 
U\delta(t-\tau),
\\
U \equiv \frac{\hbar\beta^2\mathcal{P}}{\omega_0},
\quad
\beta \equiv 2Mk_0\cos\theta.
\end{align}
The mirror position is observed via optical phase measurements using a
phase-locked loop. In the linearized regime, we can define the quantum
observation process as
\begin{align}
\hat y &= \Bk{\begin{array}{cc}
\beta & 0
\end{array}}
\Bk{\begin{array}{c}
\hat x \\ \hat p
\end{array}} + \hat z,
\\
\Avg{\hat z(t)\hat z(\tau)} &= Z\delta(t-\tau),
\quad
Z \equiv \frac{\hbar\omega_0}{4\mathcal{P}}.
\end{align}
Our linearized model is consistent with the general model of
continuous quantum non-demolition (QND) measurements
\cite{belavkin,doherty,caves}.

To apply Kalman-Bucy filtering to the estimation of
mirror position and momentum, let us define
\begin{align}
\Delta \hat x &\equiv \hat x-\est{x},
\quad
\Delta \hat p \equiv \hat p-\est{p},
\\
\bs{\Sigma} &= \Bk{\begin{array}{cc}
\avg{\Delta \hat x^2} & \frac{1}{2}\Avg{\Delta \hat x\Delta \hat p+
\Delta \hat p\Delta \hat x}
\\
\frac{1}{2}\avg{\Delta \hat x\Delta \hat p+\Delta \hat p \Delta \hat x}
& \avg{\Delta \hat p^2}
\end{array}}
.
\end{align}
In the linearized model, the Wigner distribution remains Gaussian and
non-negative provided that the initial Wigner distribution is
Gaussian, so it can be regarded as a classical phase-space probability
distribution, $\hat x(t)$, $\hat p(t)$, $\Delta\hat I(t)$, and $\hat
y(t)$ can be regarded as classical random processes with statistics
governed by the Wigner distribution, and we can apply classical
estimation theory directly. The off-diagonal components of the
variance matrix are written in terms of symmetrized operators to
ensure that they are Hermitian and also obey Wigner-distribution
statistics.

The Kalman-Bucy variance equations hence become
\begin{align}
\diff{\Sigma_{11}}{t}
&=\frac{1}{m}(\Sigma_{21}+\Sigma_{12}) -V
\Sigma_{11}^2,
\\
\diff{\Sigma_{12}}{t}
&=\frac{1}{m}\Sigma_{22}-m\omega_m^2\Sigma_{11}
-V\Sigma_{11}\Sigma_{12},
\\
\diff{\Sigma_{22}}{t}
&=
-m\omega_m^2(\Sigma_{12}+\Sigma_{21})
-V\Sigma_{12}\Sigma_{21}+U,
\\
V &\equiv \frac{4\beta^2\mathcal{P}}{\hbar\omega_0}.
\end{align}
The steady state is given by the condition $d\bs{\Sigma}/dt = 0$.
After some algebra,
\begin{align}
(\Sigma_{11})_{\rm ss}&=\frac{\hbar}{2m\omega_m}
\frac{\sqrt{2}}{Q}
\Bk{\bk{1+Q^2}^{1/2}-1}^{1/2},
\label{S11}
\\
(\Sigma_{12})_{\rm ss} &=(\Sigma_{21})_{\rm ss} = \frac{\hbar}{2}\frac{1}{Q}
\Bk{\bk{1+Q^2}^{1/2}-1},
\label{S12}
\\
(\Sigma_{22})_{\rm ss} &= \frac{\hbar m\omega_m}{2}
\frac{\sqrt{2}}{Q}
\Bk{\bk{1+Q^2}^{1/2}-1}^{1/2}\bk{1+Q^2}^{1/2},
\label{S22}
\end{align}
where
\begin{align}
Q &\equiv \frac{\sqrt{UV}}{m\omega_m^2} = 
\frac{2\beta^2\mathcal{P}}{m\omega_0\omega_m^2}
\end{align}
is a dimensionless parameter that characterizes the strength of the
measurements. The position uncertainty $\avg{\Delta \hat x^2}_{\rm ss}
= (\Sigma_{11})_{\rm ss}$ is squeezed due to the continuous QND
measurements, while the momentum uncertainty $\avg{\Delta \hat
  p^2}_{\rm ss} = (\Sigma_{22})_{\rm ss}$ is anti-squeezed due to the
radiation pressure. These results have also been derived by various
groups of people \cite{belavkin,doherty}, although here we have shown
how one can realistically implement the optical measurements of a
mechanical oscillator.

The Kalman-Bucy gain is
\begin{align}
\bs\Gamma &= \frac{4\beta\mathcal{P}}{\hbar\omega_0}
\Bk{\begin{array}{c}\Sigma_{11}\\ \Sigma_{21}
\end{array}}.
\end{align}
The estimator equation becomes
\begin{align}
\diff{}{t}\Bk{\begin{array}{c}\est{x}\\\est{p}\end{array}}  &= 
\Bk{\begin{array}{cc}-V\Sigma_{11} & 1/m\\
-V\Sigma_{21}-m\omega_m^2 & 0\end{array}}
\Bk{\begin{array}{c}\est{x}\\\est{p}\end{array}}
\nonumber\\&\quad+
\frac{V}{\beta}
\Bk{\begin{array}{c}\Sigma_{11}\\ \Sigma_{21}
\end{array}}y,
\end{align}
where $y$ is the measurement record of $\hat{y}$.  The filter
relaxation time is on the order of
\begin{align}
t_f &\sim \frac{1}{V(\Sigma_{11})_{\rm ss}} = 
\frac{1}{\sqrt{2}\omega_m[(1+Q^2)^{1/2}-1]^{1/2}},
\label{tf}
\end{align}
which decreases for increasing $Q$, so the steady state can be reached
faster for a larger $Q$. For $Q \to 0$, $t_f \to \infty$, and a steady
state does not exist.  The photon number within the filter relaxation
time is much larger than 1, and the assumption of white Gaussian
radiation pressure noise is valid, when
\begin{align}
\frac{\mathcal{P}t_f}{\hbar\omega_0} &\sim
\frac{m\omega_m}{2\sqrt{2}\beta^2\hbar}
\frac{Q}{[(1+Q^2)^{1/2}-1]^{1/2}} 
\gg 1.
\label{large_num}
\end{align}
On the other hand, the threshold constraint, which ensures that the
linearized analysis of the phase-locked loop is valid, is
\begin{align}
\beta^2(\Sigma_{11})_{\rm ss}  &= 
\frac{\beta^2\hbar}{\sqrt{2}m\omega_m}
\frac{[(1+Q^2)^{1/2}-1]^{1/2}}{Q} \ll 1.
\end{align}
This condition, apart from a factor of $4$, is the same as the
large-photon-number assumption given by Eq.~(\ref{large_num}), and
ensures that the linearized system and measurement model is
self-consistent.

The mirror position-momentum uncertainty product at steady state is
\begin{align}
\avg{\Delta \hat x^2}_{\rm ss}\avg{\Delta \hat p^2}_{\rm ss}
&=(\Sigma_{11}\Sigma_{22})_{\rm ss} \nonumber\\
&=\frac{\hbar^2}{4}
\frac{2}{Q^2}\Bk{1+Q^2-\bk{1+Q^2}^{1/2}},
\end{align}
and satisfies the Heisenberg uncertainty principle for all $Q$, as one
would expect. Furthermore, the covariances satisfy the
following relation for pure Gaussian states \cite{belavkin,doherty}:
\begin{align}
\operatorname{det}(\bs\Sigma_{\rm ss}) = 
\bk{\Sigma_{11}\Sigma_{22}-
\Sigma_{12}^2}_{\rm ss} = \frac{\hbar^2}{4},
\end{align}
indicating that the harmonic oscillator conditioned upon the real-time
measurement record becomes a pure Gaussian state at steady state.

\subsection{Smoothing errors}
From the classical estimation theory perspective, we should be able to
improve upon Kalman-Bucy filtering if we allow delay in the estimation
and apply smoothing. Here we calculate the steady-state smoothing
errors using the two-filter approach described in
Sec.~\ref{twofilter}.  The steady-state smoothing covariance matrix is
\begin{align}
\bs{\Pi}_{\rm ss} &= \bk{\bs\Sigma_{\rm ss}^{-1}+\bs\Xi_{\rm ss}^{-1}}^{-1},
\end{align}
where $\bs\Sigma_{\rm ss}$ is the steady-state forward-filter
covariance matrix, already solved and given by
Eqs.~(\ref{S11})--(\ref{S22}).  The backward-filter covariances obey
the following equations:
\begin{align}
\diff{\Xi_{11}}{t}
&=\frac{1}{m}(\Xi_{21}+\Xi_{12}) +V
\Xi_{11}^2,
\\
\diff{\Xi_{12}}{t}
&=\frac{1}{m}\Xi_{22}-m\omega_m^2\Xi_{11}
+V\Xi_{11}\Xi_{12},
\\
\diff{\Xi_{22}}{t}
&=
-m\omega_m^2(\Xi_{12}+\Xi_{21})+V\Xi_{12}\Xi_{21}-U.
\end{align}
The steady-state values for the backward filter
turn out to be almost identical to the ones for the forward filter,
\begin{align}
(\Xi_{11})_{\rm ss} &= (\Sigma_{11})_{\rm ss},
&
(\Xi_{12})_{\rm ss} &= -(\Sigma_{12})_{\rm ss},
\nonumber\\
(\Xi_{22})_{\rm ss} &= (\Sigma_{22})_{\rm ss},
\end{align}
and also satisfy the pure-Gaussian-state relation
\begin{align}
\operatorname{det}(\bs\Xi_{\rm ss}) = 
\bk{\Xi_{11}\Xi_{22}-
\Xi_{12}^2}_{\rm ss} = \frac{\hbar^2}{4}.
\end{align}
After some algebra,
\begin{align}
(\Pi_{11})_{\rm ss} &= \frac{\hbar}{8m\omega_m}
\Bk{\frac{1}{(1+iQ)^{1/2}}+\frac{1}{(1-iQ)^{1/2}}},
\label{P11}
\\
(\Pi_{12})_{\rm ss} &= 0,
\label{P12}
\\
(\Pi_{22})_{\rm ss} &= \frac{\hbar m\omega_m}{8}
\Bk{(1+iQ)^{1/2}+\bk{1-iQ}^{1/2}}.
\label{P22}
\end{align}
These results can be confirmed using the frequency-domain approach
outlined in Sec.~\ref{delay}.  The position-momentum uncertainty
product becomes
\begin{align}
(\Pi_{11})_{\rm ss}(\Pi_{22})_{\rm ss} = 
\frac{\hbar^2}{32}
\Bk{1+\frac{1}{(1+Q^2)^{1/2}}},
\label{small_uncertainty}
\end{align}
which is smaller than the Heisenberg uncertainty product
$\hbar^2/4$ by 4 to 8 times.

\subsection{\label{discuss}Discussion}
While counter-intuitive, the sub-Heisenberg uncertainties given by
Eqs.~(\ref{P11})--(\ref{small_uncertainty}) do not violate any basic
law of quantum mechanics. The reason is that we only estimate the
position and momentum of the mirror \emph{some time in the past} as if
they were classical random processes with Wigner-distribution
statistics, but it is impossible to verify our estimates by comparing
them against the mirror in the past, which has since been irreversibly
perturbed by the unknown radiation pressure noise. In classical
estimation, $x(t)$ and $p(t)$ are classical random processes unknown
to the observer but can in principle be perfectly measured or simply
decided at will by another party, so it is possible to compare one's
delayed estimates against the perfect versions and verify the
smoothing errors. In the quantum regime, however, one cannot measure
the mirror in the past more accurately without disturbing it further,
and the only way for us to obtain perfect information about the mirror
position or momentum is to perform a strong projective
measurement. Unlike the Kalman-Bucy estimates, which predict the
mirror position and momentum at present and can be verified by
performing a projective measurement at present, it is obviously
impossible to go back to the past and perform a strong projective
measurement on the mirror to verify our delayed estimates without
changing our model of the problem.

It is also impossible to perfectly reverse the dynamics of the mirror
in time and recreate the past quantum state without introducing
additional noise, because quantum-limited optical phase measurements
prevent us from obtaining any information about the optical power
fluctuations, and the dynamics of the mirror subject to the unknown
radiation pressure noise is irreversible. Thus, even though classical
estimation theory indicates that we can achieve more accurate
estimates than the Heisenberg uncertainty principle would allow,
quantum mechanics seem to forbid one from experimentally verifying the
violation. In this sense the apparent paradox is analogous to the
Einstein-Podolsky-Rosen paradox \cite{epr} and may yet have
implications for the interpretation of quantum mechanics.

In practice, while one may argue from a frequentist point of view that
delayed estimation of quantum processes is meaningless if it cannot be
verified, smoothing should still be able to improve the estimation of
a classical random process in a quantum system, such as a classical
force acting on a quantum harmonic oscillator \cite{thorne}.

\section{Conclusion}
In conclusion, we have used classical estimation theory to design
homodyne phase-locked loop for quantum optical phase estimation, and
shown that the estimation performance can be improved when delay is
permitted and smoothing is applied. We have focused on coherent
states, as it can be regarded as a classical field with additive
phase-insensitive noise upon homodyne detection, and classical
estimation techniques can be applied directly. The optimal adaptive
homodyne measurement scheme for nonclassical states remains an open
problem. Along this direction Berry and Wiseman have recently
suggested the use of Bayesian estimation for narrowband squeezed
states when the mean phase is a Wiener process
\cite{berry2}. Generalization of their scheme to more general random
processes is challenging but may be facilitated by classical nonlinear
estimation techniques \cite{vantrees,baggeroer}.

When we apply the same classical techniques to the quantum-limited
estimation of harmonic-oscillator position and momentum, we find that
the two conjugate variables can be simultaneously estimated with
inferred accuracies beyond the Heisenberg uncertainty relation, if
smoothing is performed. Although quantum mechanics seems to forbid one
from verifying the delayed estimates by destroying the evidence, this
result remains counter-intuitive and may have implications for the
interpretation of quantum mechanics. In the general context of quantum
trajectory theory \cite{carmichael}, quantum smoothing
deserves further investigation and should be useful for quantum
sensing and communication applications \cite{thorne,barnett}.

\section*{Acknowledgments}
Discussion with Howard Wiseman is gratefully acknowledged.  This work
is financially supported by the W.\ M.\ Keck Foundation Center for
Extreme Quantum Information Theory.

\end{document}